\documentclass[12pt,a4paper]{article}
\usepackage{amsmath}
\usepackage[dvips]{graphicx}
\input epsf
\usepackage{times}
\begin{document}
\begin{titlepage}
\title{Note on spin carried by quarks}
\author{ S.M. Troshin, N.E. Tyurin\\[1ex]
\small  \it Institute for High Energy Physics,\\
\small  \it Protvino, Moscow Region, 142281, Russia}
\normalsize
\date{}
\maketitle

\begin{abstract}
We note that recent results on the role of final state interactions
in deep inelastic scattering make more
complicated an extraction of  a nucleon spin fraction
carried by quarks from the
experimental data obtained in deep inelastic scattering.

\end{abstract}
\end{titlepage}
\setcounter{page}{2}
Crucial results on the essential role of final state interactions in the
 interpretation of structure functions $F_i(x,Q^2)$ in deep inelastic scattering
  have been obtained recently  \cite{brodsky}. It appeared that final state interactions
 of the struck quark with  proton remnants cannot be eliminated and this fact
  invalidates interpretation of the
 structure functions as  parton probability densities. These results lead to the important
 phenomenological implications, in particular, they provide significant
 single spin asymmetries in SIDIS \cite{bhs,burk} and Drell--Yan processes \cite{collins}.

It is evident that arguments of \cite{brodsky} can be  applied to the polarized structure
functions $G_i(x,Q^2)$ as well.  Experimentally
measured quark spin distributions \\
$\widetilde{\Delta q} (x,Q^2)$ will include then
 rescattering  effects related to the
presence of the path ordered exponent in their definitions as Fourier transform
of the matrix elements of the axial vector current operator similar
to the case of unpolarized quark distributions \cite{brodsky,collins}.
By definition\footnote{We apply definition of quark densities,
which allow straightforward  interpretations of the experimental
results;  thorough  discussion of
 definitions of parton densities has been given recently in \cite{collins1}.}
  quark spin densities $q_{\uparrow} (x,Q^2)$ and $q_{\downarrow} (x,Q^2)$
are the probability densities
 to find a quark of flavor $q$
with a given helicity value and momentum fraction in the hadron wavefunction resolved
in the transverse plane with the resolution scale determined by $Q^2$.
This allow us to interpret the first moment of the function
\[
\Delta q (x,Q^2)= q_{\uparrow} (x,Q^2)-q_{\downarrow} (x,Q^2)+
\bar q_{\uparrow} (x,Q^2)-\bar q_{\downarrow} (x,Q^2)
\]
 as  the net fractions of the nucleon
spin carried by quarks of particular flavor, and total fraction of
hadron spin carried by quarks $\Sigma$ is a sum of the first moments over all flavors.
Thus, we immediately  arrive to conclusion that
it is more difficult to  extract in a usual way quark
the spin densities
$\Delta q (x,Q^2)$
 from
the experimentally measured polarized structure functions $g_{1,2}(x,Q^2)$.
Indeed, rescattering  affects most strongly the region of small $x$ and  we can
then expect here a maximal deviation of the functions $\widetilde{\Delta q} (x,Q^2)$ from
the usual quark spin densities ${\Delta q} (x,Q^2)$. At the same time the region
 of small $x$ is most important for the calculation of the
  net spin carried by quarks and antiquarks. Since, the structure functions
  are experimentally measured in the limited region of $x$ the cancellation
  of the rescattering effects, which is valid for $\int_0^1{\Delta q} (x,Q^2)$,
  may not take place for the limited integration region.

It is well known now that the functions
  $\widetilde{\Delta q} (x,Q^2)$ and ${\Delta q} (x,Q^2)$ can be different
   due to perturbative contribution of an
  axial vector anomaly
  and this way to treat ``spin crisis'' was proposed at its early stage
  \cite{efremov,ross}.
Rescattering effects adherent
 to both polarized and
unpolarized structure functions have a different origin.

To this end, the conclusion on the increasing difficulty
of the  extraction of fraction of
spin carried by quarks from the experimental data in deep inelastic scattering due to
rescattering effects seems
inevitable and the fundamental reason for this conclusion is that we have to deal with
not free but interacting fields.  This interaction leads to strong
coherence effects  in the interaction of virtual photon with hadron.
Account for such effects can be done in the model way.
For example, modelling of these effects in the nonperturbative
 approach based on unitarity
has been performed in \cite{trtu}.

\end{document}